\documentclass[12pt]{article}

\usepackage{epsfig}

\begin{document}

\begin{center}
\baselineskip=24pt

\begin{minipage}{\linewidth}

\begin{center}
{\Large \bf Study of single muons with the Large Volume Detector at Gran 
Sasso Laboratory}
\vspace{0.5cm}

\baselineskip=18pt

\vspace{0.2cm}
{\large \bf LVD Collaboration}\\
\vspace{0.3cm}

M. Aglietta$^{14}$, E. D. Alyea$^{7}$, P. Antonioli$^{1}$,
G. Badino$^{14}$, G. Bari$^{1}$, M. Basile$^{1}$,
V. S. Berezinsky$^{9}$, F. Bersani$^{1}$, M. Bertaina$^{8}$,
R. Bertoni$^{14}$, G. Bruni$^{1}$, G. Cara Romeo$^{1}$,
C. Castagnoli$^{14}$, A. Castellina$^{14}$, A. Chiavassa$^{14}$,
J. A. Chinellato$^{3}$, L. Cifarelli$^{1 \dagger}$, F. Cindolo$^{1}$,
A. Contin$^{1}$, V. L. Dadykin$^{9}$,
L. G. Dos Santos$^{3}$, R. I.  Enikeev$^{9}$, W. Fulgione$^{14}$,
P. Galeotti$^{14}$, P. Ghia$^{14}$, P. Giusti$^{1}$, F. Gomez$^{14}$,
R. Granella$^{14}$, F. Grianti$^{1}$, V. I. Gurentsov$^{9}$,
G. Iacobucci$^{1}$, N. Inoue$^{12}$,
E. Kemp$^{3}$, F. F. Khalchukov$^{9}$, E. V. Korolkova$^{9}$
\footnote{Corresponding author, 
e-mail: korolkova@vaxmw.tower.ras.ru},
P. V. Korchaguin$^{9}$, V. B. Korchaguin$^{9}$,
V. A. Kudryavtsev$^{9 \dagger\dagger}$,
M. Luvisetto$^{1}$,
A. S. Malguin$^{9}$, T. Massam$^{1}$, N. Mengotti Silva$^{3}$,
C. Morello$^{14}$, R. Nania$^{1}$, G.Navarra$^{14}$,
L. Periale$^{14}$, A. Pesci$^{1}$, P.Picchi$^{14}$,
I. A. Pless$^{8}$, V. G. Ryasny$^{9}$,
O. G. Ryazhskaya$^{9}$, O. Saavedra$^{14}$, K. Saitoh$^{13}$,
G. Sartorelli$^{1}$,
M. Selvi$^{1}$, N. Taborgna$^{5}$, V. P. Talochkin$^{9}$,
G. C. Trinchero$^{14}$, S. Tsuji$^{10}$, A. Turtelli$^{3}$,
P. Vallania$^{14}$, S. Vernetto$^{14}$,
C. Vigorito$^{14}$, L. Votano$^{4}$, T. Wada$^{10}$,
R. Weinstein$^{6}$, M. Widgoff$^{2}$,
V. F. Yakushev$^{9}$, I. Yamamoto$^{11}$,
G. T. Zatsepin$^{9}$, A. Zichichi$^{1}$

\medskip

$^{1}$ {\it University of Bologna and INFN-Bologna, Italy}\\
$^{2}$ {\it Brown University, Providence, USA}\\
$^{3}$ {\it University of Campinas, Campinas, Brazil}\\
$^{4}$ {\it INFN-LNF, Frascati, Italy}\\
$^{5}$ {\it INFN-LNGS, Assergi, Italy}\\
$^{6}$ {\it University of Houston, Houston, USA}\\
$^{7}$ {\it Indiana University, Bloomington, USA}\\
$^{8}$ {\it Massachusetts Institute of Technology, Cambridge, USA}\\
$^{9}$ {\it Institute for Nuclear Research, Russian Academy of
Sciences, Moscow, Russia}\\
$^{10}$ {\it Okayama University, Okayama, Japan}\\
$^{11}$ {\it Okayama University of Science, Okayama, Japan}\\
$^{12}$ {\it Saitama University of Science, Saitama, Japan}\\
$^{13}$ {\it Ashikaga Institute of Technology, Ashikaga, Japan}\\
$^{14}$ {\it University of Torino and INFN-Torino, Italy} \\
\indent {\it Institute of Cosmo-Geophysics, CNR, Torino, Italy}\\
\vspace{0.2cm}
$^{\dagger}$ {\it now at the University of Salerno and INFN-Salerno, Italy}\\
$^{\dagger\dagger}$ {\it now at the University of Sheffield, UK}\\
\vspace{0.2cm}
\end{center}
\end{minipage}

\pagebreak

\begin{abstract}
The present study is based on the sample of $2.9 \times 10^{6}$ single muons 
observed by LVD at underground Gran Sasso Laboratory during 36500 live 
hours from June 1992 to February 1998. We have measured the muon intensity 
at slant depths from 3 km w.e. to 20 km w.e. Most events are 
high energy downward muons produced by meson decay in the 
atmosphere. The analysis of these muons has revealed  the 
power index $\gamma$ of $\pi$ and $K$  spectrum: $\gamma = 2.76 \pm 
 0.05$. The reminders are horizontal 
muons produced by the neutrino interactions in the rock surrounding LVD. 
The value of this flux near $90^{\circ}$ is $(6.1 \pm 2.7) \times 10^{-13}$ 
cm$^{-2}$ s$^{-1}$ sr$^{-1}$. The results are compared with 
Monte Carlo simulations and the world data.
\end{abstract}

\end{center}

\vspace{0.5cm}
\noindent Corresponding author: E.V. Korolkova, Institute for Nuclear 
Research of the Russian Academy of Science, 7a, 60th October Anniversary 
prospect, Moscow 117312, Russia

\noindent Tel: 007 095 9381930; \hspace{2cm} Fax: 007 095 9385754. 

\noindent E-mail: korolkova@vaxmw.tower.ras.ru
\pagebreak

{\large \bf 1. Introduction}
\vspace{0.3cm}

\indent The study of atmospheric muons at large depth underground is 
the subject of experimental investigations due to the following reasons. 
First, muons and muon-produced secondary particles are the 
background for underground detectors designed to search for rare 
events including the tasks of neutrino and gamma-ray astronomy. 
Second, the
calculations of atmospheric muon and neutrino fluxes are based on 
the hypothesis about primary cosmic-ray spectrum and 
hadron--hadron interactions. The 
existing deep underground detectors are not able to measure muon energy for
direct deduction of energy spectrum. But they are able to measure 
muon `depth-intensity' curve. This curve shows the  
vertical muons flux as a function of the rock (water, ice) depth and is related
to the muon propagation through the rock, muon energy spectrum at sea 
level and then, to the primary cosmic-ray 
spectrum. The connection to the muon propagation allows tests of the 
cross-sections of muon interactions with 
the rock which have been used in the program for the muon transport.

The measurements of muon intensity using the single detector and 
technique from a relatively small depths to 
the large depth where neutrino-induced muons dominate
are of special interest. Such an 
experiment observes muons at zenith angles from vertical to horizontal 
direction. The statistics for measurement of neutrino-induced flux in 
horizontal direction is small enough but the uncertainties in  
detecting
the muon direction are absent. There is no need to suppress the flux of  
atmospheric muons by a factor of about $10^{6}$ using accurate time 
measurements.
   
The LVD structure and the complicated profile of the Gran Sasso mountains
provide an opportunity to measure the muon `depth-intensity' curve for 
slant depths from 3 to 20 km w.e. and the neutrino-induced muon flux at 
horizontal direction where the atmospheric muon flux is suppressed 
due to the large slant depth. The expected number of horizontal events 
caused by neutrino-induced muons is small (about one -- two events per 
one LVD tower per year), therefore for the moment we cannot make a 
conclusion about neutrino oscillations.

In our previous paper \cite{LVD} we have presented our first results 
on the
measurement of the muon depth--intensity curve for the depth range 
of 3--20 km w.e. Since 
that time we have improved the criteria for event selection and 
increased the statistics. The analysis of Ref. \cite{LVDsp}
was based on the events with all multiplicities. Multiple muon events 
especially
for large depths are more difficult to reconstruct than single 
muons. To avoid this problem we have also performed the analysis of single 
muons using stronger criteria of run and event selection.
The analysis is based on an increased statistics compared
with our previous publications.

In Section 2 the detector, the procedure of data analysis
and conversion of the muon intensity to vertical
are briefly described. In Section 3
the results of the analysis of the 'depth -- vertical muon intensity'
relation ($I_{\mu}(x)$) are
shown.  In Section 4 we present the analysis of neutrino-induced
events.
Section 5 contains our conclusions. 

\vspace{0.5cm}
{\large \bf 2. Detector and data selection}
\vspace{0.3cm}

\indent The LVD (Large Volume Detector) has been described extensively 
elsewhere \cite{LVD,LVDe,LVDexp}. The detector is located at Gran
Sasso Laboratory, Italy. The minimal rock overburden is
3 km w.e. 
LVD consists of 3 towers. Each tower is made of 38 modules with 
dimensions 2.1 m $\times$ 6.2 m $\times$ 1.0 m. 
The data were obtained with the first LVD
tower from June, 1992, when it was put into operation, until February,
1998. Total live 
time was 36500 hours. The tower has the size of 
13 m $\times$ 6.6 m $\times$ 12 m. Each module contains eight scintillator 
counters with active volume 1.0 m $\times$ 1.5 m $\times$ 1.0 m and mass 
of liquid scintillator of 1.2 tons, and tracking detector attached to the 
bottom and one vertical side of the supporting structure. Each 
tracking detector is made of 4 layers of tubes operating in limited 
streamer mode. Each layer has independent $x$ and $y$ readout strips. These 
established the $x$ and $y$ coordinates of the hits. The tracking system 
allows the measurements of particle direction with accuracy better 
than $0.5 ^{\circ}$.

The mountain structure above Gran Sasso laboratory allows the 
measurements of muons which traversed slant depth from 3 to more than 12 km w.e.
The depths correspond to the median muon energies at sea level from 1.5 to 
40 TeV at zenith angles from $0^{\circ}$ to $90^{\circ}$. In the analysis we 
have used the sample of events containing only single muons. 
Events with all multiplicities are usually studied in the 
experiments with cosmic rays muons. Such an analysis supposes the 
accurate reconstruction of each event. The study of depth-intensity 
curve with all muon events observed by LVD  has been presented in Ref. 
\cite{LVDsp}.
The multiple muons have been considered as independent muons and 
the acceptance for both single and multiple muons has been assumed 
the same. This is a good and well proven approximation for the 
derivation of the all-particle primary spectrum. The tasks requires 
however the reconstruction of
muon events with all multiplicities and the measurements of the direction
and  slant depths with good accuracy. 
This is more difficult task for multiple muon events than for single 
ones. Most muons traverse small rock thicknesses. If the slant depth 
for a small fraction of these events is defined wrong, then the
intensity in this direction will not change much. However these 
erroneously reconstructed  events can contribute significantly to the 
muon intensity at large depth. To be sure in the precise event 
reconstruction in this analysis we have dealt with single muons only.  
The size of one LVD tower is small enough and more than  $90\%$ 
of muon events are single muons. The number of muons in bundles 
is about 10--12$\%$ of the total number of muons. Possible 
uncertainties from neglecting multiple muons are less than or 
comparable to the 
errors from including multiple-muon events with erroneous reconstruction.
In the case of single muon analysis we need to correct the 
absolute intensity  for the number of  unreconstructed events 
(multiple muons and muon induced cascades).

The trigger for muon events has been defined as follows:
{\it i)} energy deposition is grater than 30 MeV 
in at least two scintillator counters in two different modules; 
{\it ii)}  hits 
in at least three layers in any three tracking detectors (hits in at least one
layer per detector).  
The data runs have been selected as follows: runs have been
accepted if they had more than one-hour duration and a counting rate is 
within a  $15\%$ range around the mean value for the set of runs. 
Moreover we have required at least 36 from 38 tracking 
modules and 240 from 304 scintillator counters of the first tower
to be operated in any particular run. These criteria ensured the  
full and uniform acceptance of the detector.
The final muon sample after these cuts consisted of 3151580 events.
2877659 ($91\%$) events have been reconstructed as single muons. 
Multiple muons and muons accompanied by cascades constituted $9\%$. All 
reconstructed single muons were binned in two dimensional array
with a cell size of 1$^o$ in azimuthal angle $\phi$ 
and 0.01 in $\cos\theta$, where $\theta$ is the zenith angle. The accuracy 
of the reconstruction has been checked by observation of the Moon 
shadowing effect with single muon data \cite{Moon}. It is better than 
$0.65^{\circ}$.  

The acceptance for each angular bin has been calculated using the 
simulation of muons traversing LVD taking into account the detector response.
The thickness of rock crossed by the muon was determined from the 
mountain map. 

The angular distribution $N_{\mu}(\phi, \cos \theta)$ obtained in 
the experiment has been converted to the 'depth--intensity relation' 
using the formula:

\begin{equation}
I_{\mu} (x_{m}) = {{\sum_{ij} N_{\mu}(x_{m}(\phi_{j},\cos \theta_{i}))}
\over {\sum_{ij}(S(x_{m}(\phi_{j},\cos \theta_{i})) \times 
\epsilon(x_{m}(\phi_{j},\cos \theta_{i}))  \times
\Omega_{ij} \times T )}}
\end{equation}

\noindent where the summing up has been done over all angular bins $(\phi_{j},
\cos \theta_{i})$ contributing to the depth $x_{m}$; 
$S(x_{m}(\phi_{j},\cos \theta_{i}))$ is the cross-section of the 
detector in the plane perpendicular to the muon track at the angle 
$(\phi_{j}, \cos \theta_{i})$; $\epsilon(x_{m}(\phi_{j},\cos \theta_{i}))$  
is the efficiency of muon detection and reconstruction; $\Omega_{ij}$ is 
the solid angle for the angular bin; $T$ is the live time. Angular bins 
with $\epsilon$ less than 0.03 were excluded from the analysis. The 
acceptance $A$ is defined as follows

\begin{equation}
A(\cos\theta,\phi) =\epsilon (\cos \theta,\phi) 
\times \Omega(\cos \theta,\phi) \times S(\cos \theta,\phi)
\end{equation}

\noindent and is shown in Figure 1.

The intensity of muons at zenith angle $\theta$ was assumed to 
be related to the vertical intensity $I_{0}$
through the relation

\begin{equation}
I(x_{m}, \theta) = I_{0}(x_{m}, \theta=0^{\circ}) / \cos 
\theta^{\star}_{i}
\end{equation}

\noindent where

\begin{equation}
\cos \theta^{\star}_{i} = {{I^{c}_{\mu}(x_{m},\cos \theta = 1)} \over {I^{c}_{\mu}
(x_{m},\cos \theta_{i})}}  
\end{equation} 

\noindent is the ratio of calculated muon intensity at 
$\cos \theta = 1$ to that at  $\cos \theta_{i}$. This relation is 
valid for muons of atmospheric origin if we neglect the  
contribution of the prompt muons from charmed particles. According to 
the LVD data the ratio of prompt muons to pions does not exceed $2 
\times 10^{-3}$ at 95 $\%$ confidence level \cite{prompt}. 
 
For depth -- intensity relation the bin width of 200 m. w. e. has 
been chosen. For depth more than 9 km w.e. we have chosen bins 
with the width of 
500 km w.e. to increase the statistics for each bin. The 
conversion of muon intensity to the middle points of each depth bin 
has been done using formula:
 
\begin{equation}
I_{\mu}^m(x_i)=I_{\mu}^m(x_m) {{I_{\mu}^c(x_i)}\over{I_{\mu}^c(x_m)}}
\end{equation}

\noindent where $I_{\mu}^m(x_m)$ and $I_{\mu}^c(x_m)$ are the measured and calculated
muon intensities at the weighted average depth $x_m$ which 
corresponds to the depth bin with the middle value of $x_i$;
$I_{\mu}^m(x_i)$ and $I_{\mu}^c(x_i)$ are the derived and calculated
muon intensities at the depth $x_i$ which is the middle point of the
depth bin. The values of $x_m$ have been obtained by averaging the
depths for all angular bins contributing to the given depth bin
with a weight equal to the detected number of muons. To calculate
the muon intensities at $x_m$ and $x_i$ we have used the muon spectrum 
at sea level with previously estimated parameters
\cite{LVD,LVDsp} (see also equation \ref{Gaisser spectrum}) 
and the simulated muon
survival probabilities.
Since the width of depth bins is quite small (200 m w.e. for depth bins
with high statistics) and the number of angular bins
contributing to each depth bin is quite large (several hundreds), 
the conversion factor does not exceed 10\%.

\vspace{0.5cm}
{\large \bf 3. `Depth -- vertical intensity' relation in Gran Sasso
rock} 
\vspace{0.3cm}

\indent To calculate the intensity of muons underground requires the 
intensity of muons at the surface as a function of energy and zenith angle, 
and the survival probability as a function  of slant depth of rock 
traversed:

\begin{equation}
I_{\mu}(x,\cos\theta)=\int_0^{\infty} P(E_{\mu 0},x)
{{d I_{\mu 0}(E_{\mu 0},\cos\theta)} \over {d E_{\mu 0}}}
d E_{\mu 0} \label{muon intensity}
\end{equation}

\noindent where $P(E_{\mu 0},x)$ is the  probability of muon with an
initial energy $E_{\mu 0}$ at sea level to reach the depth $x$
and ${{d I_{\mu 0}(E_{\mu 0},\cos\theta)} / {d E_{\mu 0}}}$
is the muon spectrum at sea level at zenith angle
$\theta$. The intensity at the surface in the units of (cm$^{2}$ s 
sr GeV)$^{-1}$ can be approximated by \cite{Gaisser}:

\begin{eqnarray}
{{d I_{\mu 0} (E_{\mu 0}, \cos\theta)}\over{d E_{\mu 0}}}
& = &
A \times 0.14 \times E_{\mu 0}^{-\gamma} \nonumber \\
& \times &
\left({{1}\over{1+{{1.1 E_{\mu 0} \cos\theta^{\star}}\over{115 GeV}}}}+
{{0.054}\over{1+{{1.1 E_{\mu 0} \cos\theta^{\star}}\over{850 GeV}}}}
\right)
\label{Gaisser spectrum}
\end{eqnarray}

\noindent where the values of $cos\theta$ have been substituted by
$cos\theta^{\star}$ which have been taken from Ref. \cite{Volkovacos}.
According to Ref.\cite{Volkovacos} $cos\theta^{\star}=
E_{\pi,K}^{cr}(cos\theta=1)/E_{\pi,K}^{cr}(cos\theta)$, where
$E_{\pi,K}^{cr}$ are the critical energies of pions and kaons.
Equation \ref{Gaisser spectrum} has been obtained under a simple assumption 
of scaling in the high-energy hadron-nucleus interactions.
Under this assumption the power index of primary spectrum,
$\gamma$, is expected to be equal to that of meson (pion + kaon) spectrum,
$\gamma_{\pi,K}$. 

The muons were tracked trough the rock using propagation code MUSIC 
\cite{MUSIC} to calculate the muon survival probabilities  $P(E_{\mu 
0},x)$. The 
stochasticity of all processes of muon interaction with matter 
(nuclear interaction, pair production, bremsstrahlung and ionization)
has been taken into account. The 
cross-sections were taken from Ref. \cite{BBn,KP,KKP}. The muon 
intensities calculated with bremsstrahlung cross-section from 
Ref. \cite{KKP} 
are lower than those with bremsstrahlung cross-section 
from Ref. \cite{BBb} which was used in our previous paper \cite{LVD}. 
Using of the cross-section from  Ref. \cite{BBb} will result in a higher 
power index (softer muon spectrum) compared to the  cross-section from 
Ref \cite{KKP}. The difference in power index is of order of 0.01. 

The measured `depth -- intensity' curve is shown in Figure 2 together 
with the best fit. 
The underground muon flux observed at a slant depth $x$ and zenith 
angle $\theta$ has 
the two-component nature  and can be presented as:

\begin{equation}
I_{\mu}(x, \theta) = I_{\mu}^{(\mu)}(x, \theta) + I_{\mu}^{(\nu)}(\theta)
\end{equation}

\noindent where $I_{\mu}^{(\mu)}(x, \theta)$ is the contribution 
of atmospheric muons and $I_{\mu}^{(\nu)}(\theta)$ 
denotes the contribution of muons from neutrino 
interactions in the rock surrounding detector. For slant depths of 
13--20 km w.e. the muons seen in LVD are of the 
later origin. The last experimental point in Figure 2 
corresponds to the 
neutrino-induced muon flux. This flux was measured at the depth of 13-20 km 
w.e. To convert the flux of neutrino-induced muons to vertical 
intensity we used the calculated ratio of horizontal and vertical 
fluxes of neutrino-induced muons which is equal to 2.1 at energy 
threshold of 1 GeV for most models of atmospheric neutrino production:

\begin{equation}
I_{\mu}^{\nu}(x, \theta =0^{\circ}) = I_{\mu}^{\nu}(x, \theta 
=90^{\circ})/2.1
\end{equation}

The `depth -- intensity' curve has been fitted with the
calculated function with two
free parameters: additional normalisation constant, $A$, and
the power index of atmospheric pion and kaon spectrum, $\gamma$.
As a result of the fitting procedure the following values
of the free parameters have been obtained:
$A=1.59 \pm 0.50$, $\gamma=2.76 \pm 0.05$ for muon energies at sea 
level from 1.5 to 40 TeV. The errors of the parameters include both
statistical and systematic uncertainties. The latter one takes into account
possible uncertainties in the depth, rock composition, density and the 
uncertainty in the cross-sections used
to simulate muon transport through the rock. These values are
in good agreement with the results of similar analysis 
performed for muon events with all multiplicities 
observed by the first LVD tower during 
21804 hours of live time: 
$A=1.95 \pm 0.50$, $\gamma=2.78 \pm 0.05$. 
Note that the estimates of the parameters $A$ and $\gamma$ are strongly
correlated. The larger the value of $\gamma$ is, the larger
the normalisation factor $A$ should be.

We have repeated the fitting procedure for restricted depth ranges.
The results of this test are presented in Table 1.  
The results show that the power index is the same within errors 
for all depth ranges.

Neutrino-induced muon flux has not been included in fit procedure but 
has been added to 
the best fit at $2.5 \times 10^{-13}$ cm$^{-2}$s$^{-1}$sr$^{-1}$ level. 
Dashed curves in Figure 2 show possible values of muon intensities  
if we take into account uncertainties in the  calculation of atmospheric 
neutrino spectrum at sea level, structure functions, corrections for 
quasielastic scattering and energy threshold of detector. The 
experimental value   $(2.9 \pm 1.3) \times 10^{-13}$ 
cm$^{-2}$s$^{-1}$sr$^{-1}$  
is in agreement with the calculated one 
$(2.5 \pm 0.5) \times 10^{-13}$ cm$^{-2}$s$^{-1}$sr$^{-1}$.
It also agrees with the compiled world results on underground muon
intensities  presented by Crouch in  Ref.\cite{78} where 
the flux of neutrino-induced muons is equal to
$(2.17 \pm 0.21)  \times 10^{-13}$ cm$^{-2}$
s$^{-1}$sr$^{-1}$.

If the formula from  Ref.\cite{Volkovasp} is used for the muon
spectrum at sea level instead of eq. (\ref{Gaisser spectrum}), the
best fit value of $\gamma$ will be decreased by 0.04-0.05.

The value of $\gamma$ obtained with
LVD data is in reasonable agreement with the results of other
surface and underground experiments: DEIS
\cite{DEIS}, MUTRON \cite{MUTRON}, MIPhI \cite{MIPhI} 
(the energies of these experiments 
correspond to first few points of our depth-intensity curve),  ASD 
\cite{ASD}, NUSEX \cite{NUSEX},
MACRO \cite{MACROc}, MSU \cite{MSU} (if we consider the muon spectrum 
from \cite{Volkovasp} in the latter case).
LVD data disagree with results of Baksan 
Scintillator Telescope and KGF \cite{Baksan, KGF}. The 
difference here is likely due to the different methods of 
measurements, the applied analysis procedure in each experiment and 
uncertainties in the knowledge of overburden composition.
  
\vspace{0.5cm}
{\large \bf 4. Neutrino-induced muons. }
\vspace{0.3cm}

\indent Let us describe the evaluation of horizontal neutrino-induced 
muon flux in more details. High energy neutrinos will 
produce high energy muons in the rock. These muons will have enough 
energy to traverse the entire detector. Reconstructed muons traversed
rock thickness  greater than 12 km w.e. have been considered as candidates for 
neutrino-induced muons. These depths  correspond to the zenith 
angles more than $85^{\circ}$. We have recorded 95 such candidates during 36500 
hours of LVD life time. A careful visual scan of all these tracks   
eliminated 5 candidates from the sample because of 
confusion in the pattern recognition.

Since the timing of the 
LVD experiment (12.5 ns) is not sufficient to determine the direction of a 
track crossing one tower there is a two-fold ambiguity in the direction 
for each measured 
track. In other words LVD cannot discriminate between muon direction
$(\theta, \phi)$ and $(180^{\circ}-\theta, 180^{\circ}+\phi)$. For 
$\theta < 90^{\circ}$it is reasonable to assume that muons come from 
above, since the rock thickness above the horizon is smaller.
 Gran Sasso mountain has a very complicated profile 
and for many bins at $\theta \approx 90^{\circ}$ with $x > 12$ km w.e. 
the slant depth for inverse 
direction  $x_{1}$ is appear to be less than 8 km w. e. The muon 
intensity for 8 km w.e. is 80 times greater than the intensity for 12 
km w.e. In this case we assume that the muon came from the 
direction with smaller slant depth. Near horizontal muons with reconstructed slant 
depth greater 12 km w.e. and slant depth in opposite direction less 
than 8 km w.e. were excluded from neutrino-induced candidates and 
considered as bins $(\theta, 180^{\circ}+\phi)$. Totally we had 67 such 
events.

Some angular bins with slant depth greater than 12 km w.e. are 
surrounded by bins with smaller slant depths. According to the 
calculations of Ref. \cite{MUSIC} the average angular 
deviation of muons is $0.45^{\circ}$ at 10 km w.e. and it is caused 
mainly by multiple Coulomb scattering. The probability of muon 
coming from the direction with smaller slant depth is greater. We have
considered such a muon as coming from the direction with smaller depth 
assuming that it had been
recorded into the bin with greater depth due to the reconstruction 
error or scattering.

Five muons produced in neutrino interactions 
with surrounding rock have survived all cuts for slant depths $x >$ 13 km w. e.,  
for  $x >$ 14 km w. e.  we have found four such events and 
for $x >$ 15 km w.e. there are two neutrino--induced muons.

A Monte Carlo has been used to estimate the expected number of 
neutrino-induced muons. The spectrum of neutrino-induced muons
has been calculated following the formula:

\begin{equation}
{{d N_{\mu}} \over {d E_{\mu}}} =\int_{E_{\mu}}^{\infty} {{d N_{\nu}} \over 
{d E_{\nu}}} \times 
{{P(E_{\nu},E_{\mu}}) \over {d E_{\mu }}}
d E_{\nu} \label{neutrino induced intensity}
\end{equation}

\noindent $d N_{\nu}/ d E_{\nu}$ represents the neutrino spectrum, 
$P(E_{\nu},E_{\mu}) /d E_{\mu }$ is the probability that a neutrino 
produces a muon in the interval $(E_{\mu}, E_{\mu}+dE_{\mu})$. 
We have used the Bartol neutrino flux \cite{AGLS}, which has a systematic
uncertainty $\pm 14 \%$, and the Morfin and Tung \cite{MT}, as well 
as Duke and Owens \cite{DO} parton distributions functions, 
which result in less than 1$\%$ difference in muon spectra.
The major sources of uncertainties in the 
neutrino-induced muon flux are the uncertainty in the neutrino 
fluxes and neutrino cross-sections because 
of required extrapolations of the structure functions to small $x << 
10^{-4}$. For neutrino-induced muons 
the calculations of atmospheric neutrino flux by various 
authors differ by 
as much as 17\%. Different standard parameterisation of charged 
current cross sections also differ by as much as 13\% \cite{GHS}).         
 
Table 2 shows the number of muons observed by LVD during 36500 
hours as well as calculated values 
(the uncertainty of calculations is 20\%).            

The observed number of muons at large slant depth  
agrees with predictions  within errors. The measured neutrino induced 
horizontal muon flux is 
$(6.1 \pm 2.7) \times 10^{-13}$ cm$^{-2}$ s$^{-1}$ sr$^{-1}$,
while the calculated one is $(5.2 \pm 1.1) \times 10^{-13}$
cm$^{-2}$ s$^{-1}$ sr$^{-1}$. Our measured value agrees with the 
results of other experiments:
Soudan-2 $(5.00 \pm 0.55 \pm 0.51) \times 10^{-13}$ 
cm$^{-2}$ s$^{-1}$ sr$^{-1}$  \cite{Soudan}, Frejus 
(the flux recalculated for our energy threshold)
$(4.77 \pm 0.86) \times 10^{-13}$ cm$^{-2}$ s$^{-1}$ sr$^{-1}$ \cite{Frneu},
in South Africa mine $(4.59 \pm 0.42) \times 10^{-13}$
cm$^{-2}$ s$^{-1}$ sr$^{-1}$ \cite{SA}. 

\vspace{0.5cm}
{\large \bf 5. Conclusions}
\vspace{0.3cm}

\indent  We have measured the underground muon intensity as a 
function of the slant depth in the range of 
3 -- 20 km w.e. The analysis of `depth--intensity` relation 
in the depth range 3 -- 12 km w.e. has been done to 
obtain the power index of differential energy spectrum 
of the pions and kaons in the atmosphere
$\gamma=2.76 \pm 0.05$  in the energy range 
of 1 - 40 TeV. The errors include both statistical and 
systematic uncertainties
with the systematic error due to the uncertainty of the muon
interaction cross-sections dominating. Our results are in 
good agreement with other experiments. 
Muons traversed slant depth more than 13 km w.e. were analysed to
obtain the horizontal flux of neutrino-induced muons. This flux is 
equal to $(6.1 \pm 2.7) \times 10^{-13}$ cm$^{-2}$ s$^{-1}$ sr$^{-1}$ 
and is consistent 
with our calculations and reported results of  other experiments.
Our fit to this data at the `depth--intensity' curve is in a good 
agreement with the fit of Crouch, which is a summary of various experiments.

\vspace{0.5cm}
{\large \bf Acknowledgments}
\vspace{0.3cm}

\indent The Collaboration wishes to thank the staff of Gran Sasso 
Laboratory for assistance. This work is supported by the 
Russian Ministry of Industry, Science and Technologies,
the Italian Institute for Nuclear Phy\-sics and the
Russian Found for Basic Research  (grant 00-02-16112).

\pagebreak

\begin{table}[htb]
\caption{ The value of power index of meson spectrum for various 
depth ranges. The errors are statistical only.}
\vspace{1cm}

\begin{center}
\begin{tabular}{|c|c|c|}\hline
Depth interval, km w.e. &
$\gamma$ &
$\chi^{2}$/d.o.f. \\
\hline
3 -- 12 & 2.76 $\pm$ 0.02 & 25.8/34 \\
4 -- 12 & 2.78 $\pm$ 0.03 & 19.0/29 \\
5 -- 12 & 2.79 $\pm$ 0.04 & 16.8/24 \\
6 -- 12 & 2.82 $\pm$ 0.06 & 14.8/19 \\
7 -- 12 & 2.94 $\pm$ 0.14 & 11.4/14 \\
8 -- 12 & 2.76 $\pm$ 0.22 & 6.7/9 \\
9 -- 12 & 2.60 $\pm$ 0.50 & 3.3/4 \\
\hline
\end{tabular}  
\end{center}

\end{table}

 
\begin{table}[htb]
\caption{ The number of muons at large depths observed by LVD ($N_{obs}$)
and calculated values ($N_{\mu}^{atm}$ -atmospheric muons, 
$N_{\mu}^{\nu}$ - neutrino induced muons, 
$N_{\mu}^{tot}$ =$N_{\mu}^{atm}$ + $N_{\mu}^{atm}$).}
\vspace{1cm}

\begin{center}
\begin{tabular}{|c|c|c|c|c|}\hline
$h$, km.w.e. & $N_{\mu}^{atm}$ &
$N_{\mu}^{\nu}$ &
$N_{\mu}^{tot}$ &
$N_{obs}$\\
\hline
$h > 13$ km.w.e. & 0.45 & 3.75 & 4.20 & 5 \\
$h > 14$ km.w.e. & 0.31 & 2.92 & 3.23 & 4 \\
$h > 15$ km.w.e. & 0.03 & 1.50 & 1.53 & 2 \\
\hline
\end{tabular}  
\end{center}

\end{table}


\begin{figure}[htb]
\begin{center}
\epsfig{figure=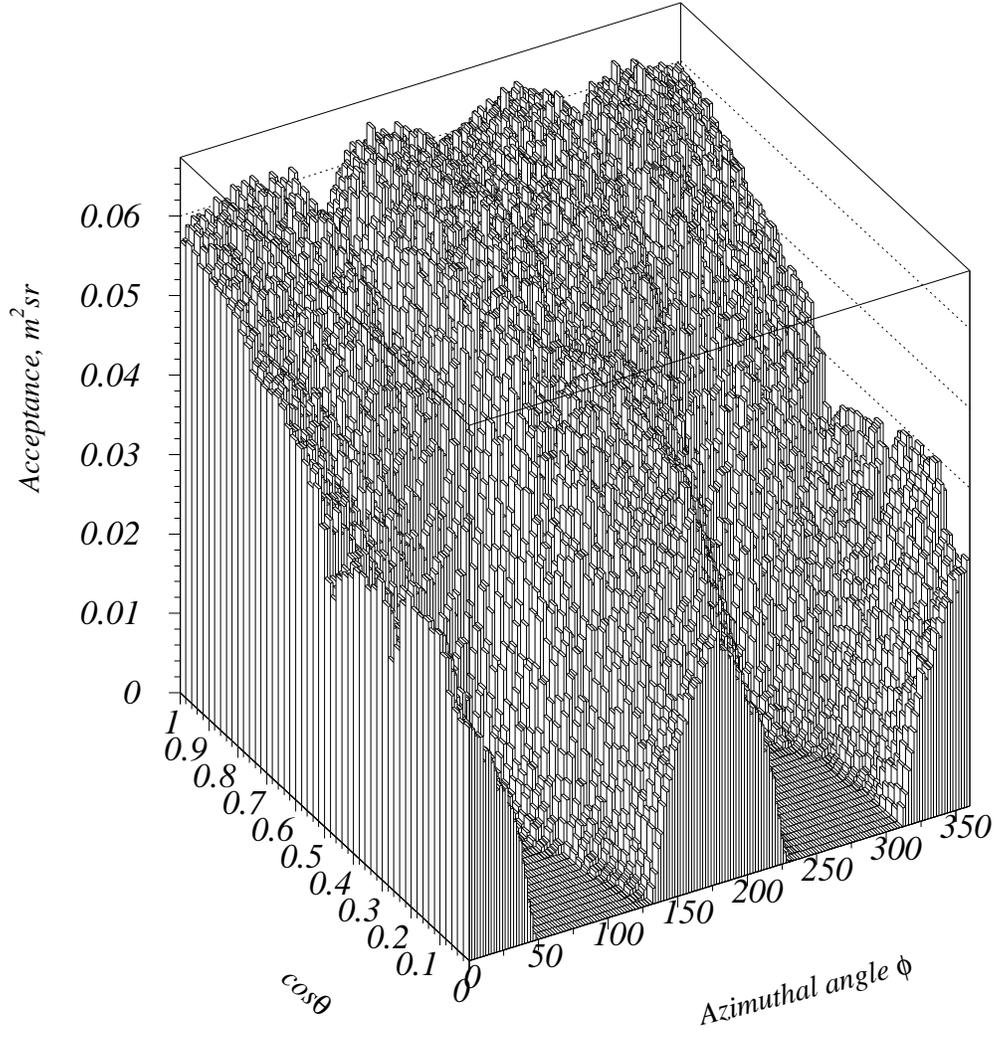,width=15cm}
\caption{ The LVD acceptance for single muons as a function of the 
cosine of the zenith angle $\theta$ and azimuthal angle $\phi$. The
angular cell for this plot was chosen as 0.02 ($\cos \theta$) 
$\times$ 2$^{\circ}$ ($\phi$).}
\end{center}
\end{figure}


\begin{figure}[htb]
\begin{center}
\epsfig{figure=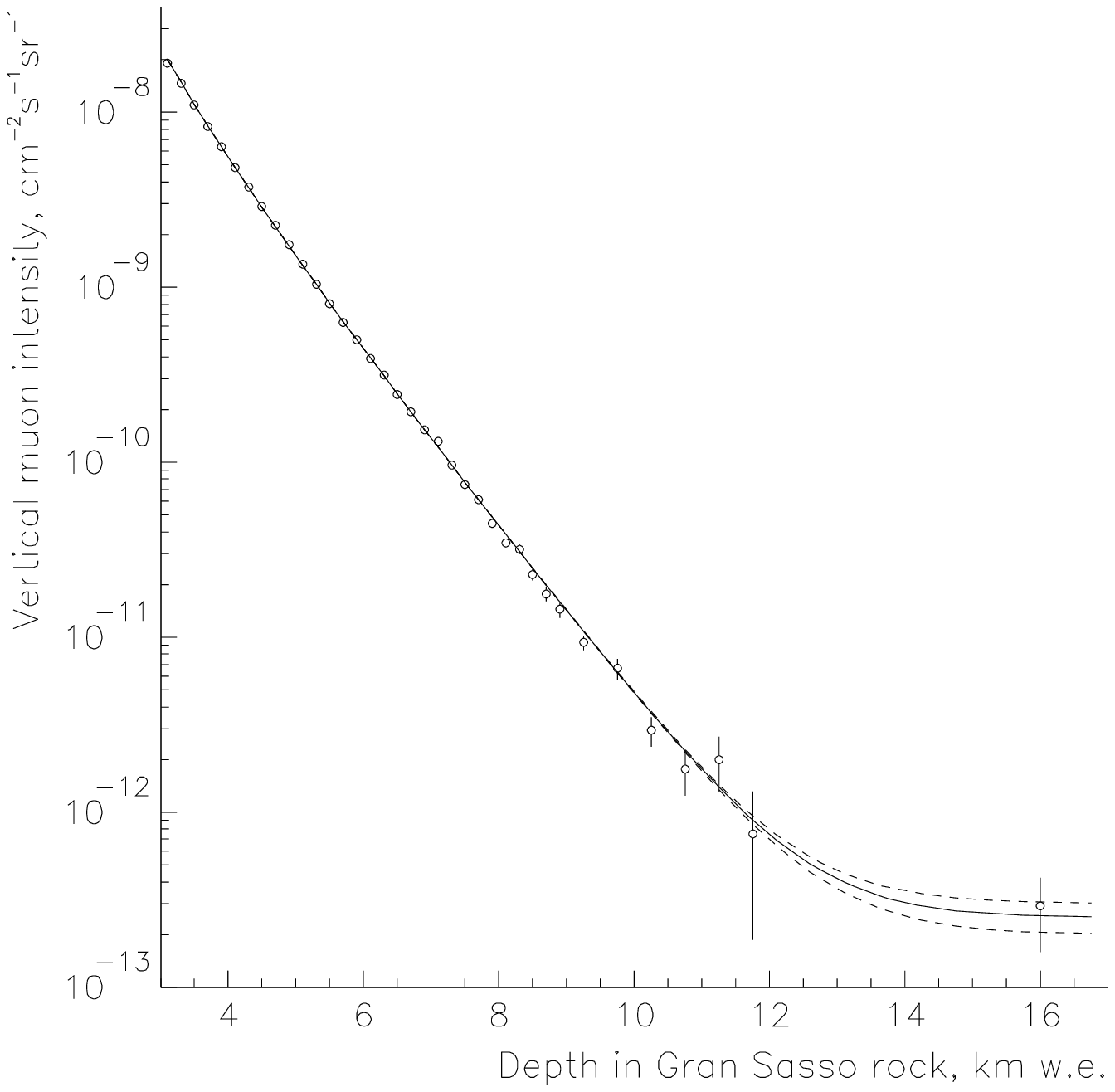,width=15cm}
\caption{ 'Depth -- vertical muon intensity' relation in Gran Sasso rock.
LVD data are presented together with the best fit (solid curve).
Dashed curves show the calculated intensities for maximal and minimal 
contributions from neutrino-induced muons (see text for details).}
\end{center}
\end{figure}


\vspace{0.5cm}
\begin{thebibliography}{99}
\bibitem{LVD}
LVD Collabotarion (M. Aglietta et al.),
{\em Astroparticle Phys.\/} {\bf 3} (1995) 311.


\bibitem{LVDsp}
LVD Collaboration (M. Aglietta et al.),
{\em Phys. Rev. D\/} {\bf 58} (1998) 092005. 

\bibitem{LVDe}
LVD Collaboration (G. Bari et al.),
{\em Nuclear Instruments and Methods in Physics Research.\/} {\bf A264} 
(1988) 5.

\bibitem{LVDexp}
LVD Collaboration (G. Bari et al.),
{\em Nuclear Instruments and Methods in Physics Research.\/} {\bf A277} 
(1989) 11.

\bibitem{Moon}
LVD Collaboration (M. Aglietta et al.), presented by E.Korolkova ,
{\em Proc. 26 Intern. Cosmic Ray Conf.\/} (Salt Lake city) {\bf 7}
(1999) 218. 

\bibitem{prompt}
LVD Collaboration (M. Aglietta et al.),
{\em Phys. Rev. D\/} {\bf 60} (1999) 112001. 

\bibitem{Gaisser}
T. K. Gaisser,
{\em Cosmic Rays and Particle Physics\/}
(Cambridge University Press, 1990)

\bibitem{Volkovacos}
L. V. Volkova,
{\em Preprint Lebedev Physical Institute N 72\/} (1969).

\bibitem{MUSIC}
P. Antonioli et al.,
{\em Astroparticle Phys.\/} {\bf 7} (1997) 357.

\bibitem{BBn}
L. B. Bezrukov and E. V. Bugaev,
{\em Proc. 17th Intern. Cosmic Ray Conf.\/} (Paris) {\bf 7} (1981) 90.

\bibitem{KP}
R. P. Kokoulin and A. A. Petrukhin,
{\em Proc. 12th Intern. Cosmic Ray Conf.\/} (Hobart) {\bf 6} (1971)
2436.

\bibitem{KKP}
S. R. Kelner, R. P. Kokoulin, and A. A. Petrukhin,
{\em Physics of Atomic Nuclei\/} {\bf 60} (1997) 576.

\bibitem{BBb}
L. B. Bezrukov and E. V. Bugaev,
{\em Proc. 17th Intern. Cosmic Ray Conf.\/} (Paris) {\bf 7} (1981) 102.

\bibitem{78}
M. Crouch,
{\em Proc. 20th ICRC\/} (Moscow) {\bf 6} (1987) 165. 

\bibitem{Volkovasp}
L. V. Volkova, G. T. Zatsepin, and L. A. Kuzmichev,
{\em Sov. J. Nucl. Phys.\/} {\bf 29} (1979) 1252.

\bibitem{DEIS}
O. C. Allkofer et al.,
{\em Proc. 17th Intern. Cosmic Ray Conf.\/} (Paris) {\bf 10} (1981) 321.

\bibitem{MUTRON}
S. Matsuno et al.,
{\em Phys. Rev. D\/} {\bf 29} (1984) 1.

\bibitem{MIPhI}
V. D. Ashitkov et al.,
{\em Proc. 19th Intern. Cosmic Ray Conf.\/} (La Jolla) {\bf 8} (1985)
77.

\bibitem{ASD}
F. F. Khalchukov et al.,
{\em Proc. 19th Intern. Cosmic Ray Conf.\/} (La Jolla) {\bf 8} (1985)
12;
R. I .Enikeev et al.,
{\em Sov. J. Nucl. Phys.\/} {\bf 47} (1988) 1044.

\bibitem{NUSEX}
G. Battistoni et al.,
{\em Nuovo Cimento\/} {\bf 9C} (1986) 196.

\bibitem{MACROc}
M. Ambrosio et al. (MACRO Collaboration),
{\em Phys. Rev. D\/} {\bf 52} (1995) 3793.

\bibitem{MSU}
N. P. Il'ina et al.,
{\em Proc. 24st Intern. Cosmic Ray Conf.\/} (Rome) {\bf 1}
(1995) 524.
 
 \bibitem{Baksan}
Yu. M. Andreyev, V. I. Gurentsov, and I. M. Kogai,
{\em Proc. 20th Intern. Cosmic Ray Conf.\/} (Moscow) {\bf 6} (1987) 200.

\bibitem{KGF}
M. R. Krishnaswami et al.,
{\em Proc. 18th Intern. Cosmic Ray Conf.\/} (Bangalore) {\bf 11}
(1983) 450.

\bibitem{AGLS}
V. Agraval, T. K. Gaisser, P. Lipari, T. Stanev, 
{\em Phys. Rev. D\/} {\bf 53} (1996) 1314.

\bibitem{MT}
J. G. Morfin, W. K. Tang
{\em Zeitchift fur Physics C\/} {\bf 52} (1991) 13.

\bibitem{DO}
D. W. Duke, J. F. Owens,
{\em Phys. Rev. D\/} {\bf 30} (1984) 49.

\bibitem{GHS}
T. Gaisser, F. Halzen, T. Stanev,
{\em Phys. Reports\/} {\bf 258} (1995) 173.

\bibitem{Soudan}
D. Demuth and M. Goodman,
{\em Proc. 27th ICRC\/} (Hamburg) {\bf HE} (2001) 1090.

\bibitem{Frneu}
W. Rhode et al.,
{\em Astroparticle physics\/} {\bf 4} (1996) 217.

\bibitem{SA}
M. Crouch et al.,
{\em Phys. Rev. \/} {\bf 18} (1978) 2239. 

\end{thebibliography}
\end{document}